\documentclass[paper]{JHEP3}
\usepackage{epsfig,cite}

\def    \be             {\begin{equation}}
\def    \ee             {\end{equation}}
\def    \ba             {\begin{eqnarray}}
\def    \ea             {\end{eqnarray}}

\def    \=              {\;=\;}
\def    \frac           #1#2{{#1 \over #2}}

\def \ss{\scriptstyle}

\title{Rise and Fall of the Bottom Quark Production Excess
\footnote{Talk given
at the La Thuile (February 2004), Moriond QCD (March 2004), IFAE (April 2004)
and Loops and Legs (April 2004) conferences. To be published, 
in slightly modified forms, in their respective Proceedings.}
}
                                                     
\author{Matteo Cacciari \\
LPTHE, Universit\'e Pierre et Marie Curie (Paris 6), France\\
  }

\abstract{  
We review the history of comparisons between bottom production
measurements and QCD predictions. We challenge the
existence of a `significant discrepancy', and argue that standard
approaches to QCD calculations do a good job in describing the experimental
findings. }

\preprint{LPTHE-P04-03}

\begin{document}

\section{Introduction}

For the past ten years or so, a rumour has been making the rounds of particle
physics Conferences and Workshops: measured bottom quark production rates are
significantly larger than predicted by next-to-leading order QCD. 

Such a state of affairs casts doubts on the ability of QCD
to properly describe this
process, and opens the way to speculations that either
drastic improvements to QCD calculations are needed, or even that
effects beyond the Standard Model might be at work.

We shall challenge the rumour under two main aspects, taking aim
especially at the words ``measured'' and ``significantly''. It is our
hope that, once done, we shall have convinced the reader that
standard approaches to QCD calculations
do a good job in describing the experimental findings.

Let us first briefly review the history of the experimental results.
Measurements of the bottom transverse momentum ($p_T$) spectrum at
colliders began in the late
80's, when the UA1 Collaboration, taking data at the CERN $Sp\bar{p}S$
with  $\sqrt{S} =$~546 and 630 GeV, published results  for the  $p_T > m_b$
(the bottom quark mass) region~\cite{Albajar:1986iu,Albajar:1990zu}. These
results were compared to the then recently completed next-to-leading order
(NLO), i.e. order $\alpha_s^3$, 
calculation~\cite{Nason:1987xz,Nason:1989zy,Beenakker:1988bq,Beenakker:1990ma}, 
and were found to be in good agreement.

During the 90's the 
CDF~\cite{Abe:1992fc,Abe:1993sj,Abe:1993hr,Abe:1994qk,Abe:1995dv,Acosta:2001rz,Acosta:2002qk} 
and D0~\cite{Abachi:1994kj,Abachi:1996jq,Abbott:1999wu,Abbott:1999se}
Collaborations also measured the bottom quark $p_T$ distribution in $p\bar p$
collisions at the Fermilab Tevatron at $\sqrt{S} = 1800$~GeV. Apparently at
odds with the UA1 results, the Tevatron data seemed to display an 
excess with respect to NLO QCD predictions.  At the same time, rates for bottom
production that appeared higher than QCD predictions were also 
observed   in $\gamma\gamma$ collisions by
three~\cite{Acciarri:2000kd,opal-gg,delphi-gg} LEP experiments, and
by the H1~\cite{Adloff:1999nr} and ZEUS~\cite{Breitweg:2000nz} Collaborations
in $ep$ collisions at HERA.

By the end of the millennium the `excess' was therefore apparently so firmly
established that the experimental papers showed no shyness in proclaiming it to
the world, e.g. ``{\it The differential cross section is measured to be $2.9 \pm 0.2\; (\mathrm{stat}\oplus\mathrm{syst}_{p_T}) \pm 0.4\;
(\mathrm{syst}_{fc})$ times
higher than the NLO QCD predictions...}''~\cite{Acosta:2001rz} and ``{\it
...NLO QCD calculations underestimate $b$ quark production by a factor of four
in the forward rapidity region}''~\cite{Abbott:1999wu}.

Despite this seemingly overwhelming evidence, we shall argue that QCD is
instead rather successful in predicting bottom production rates. 
Improved theoretical
analyses~\cite{Cacciari:2002pa,Cacciari:2003uh} and more recent
experimental measurements by the CDF~\cite{cdfrun2} and
ZEUS~\cite{Chekanov:2003si} Collaborations support this claim, which is
also borne out by a critical reconsideration of previous results.

\section{The Paradigm}
\label{sec:paradigm}

We shall take NLO QCD calculations as a benchmark for comparisons. We shall
require the experimental measurements to be genuine observable quantities. By
this we mean that as a matter of principle we do not wish to compare `data'
for, e.g., $b$-quark $p_T$ distributions, since such a quantity is clearly an
unphysical one: the quark not being directly observed, its cross sections have
to be inferred rather than directly measured.

A meaningful comparison will therefore be one between a physical cross section
and a QCD calculation with at least NLO accuracy. Non-perturbative information,
where needed, will have to be introduced in a minimal and self-consistent way.
This means that we refrain from using unjustified models, and we shall only
include non-perturbative information that has been extracted from one
experiment and then employed in predicting another observable, using
the same underlying perturbative framework in both cases. Such a
precaution allows for a good matching between the perturbative and the
non-perturbative phases, a necessity in that only the combination of the two
steps leads to an unambiguous measurable quantity.

In practice, the non-perturbative information relative to the
hadronization of the $b$-quarks into $B$-hadrons is extracted from LEP
data with a calculation~\cite{Mele:1990cw} which has NLO + NLL
accuracy\footnote{By NLL accuracy we mean that large terms of
quasi-collinear origin, proportional to  $\alpha_s^n\log^n(Q^2/m_b^2)$
and  $\alpha_s^n\log^{n-1}(Q^2/m_b^2)$, $Q$ being the LEP
centre-of-mass energy $\sim~91.2$ GeV, are resummed to all orders.}.
The framework presented in~\cite{Cacciari:2002pa} is used: the LEP (or
SLD) data~\cite{Heister:2001jg,Abe:2002iq,Abbiendi:2002vt,delphi} are
translated to  Mellin moments space, and only the moments around $N=5$
are fitted. This ensures that it is the {\sl relevant} part of the
non-perturbative information which is properly 
determined.\footnote{One readily
realizes\cite{Frixione:1997ma,Nason:1999ta}  that these moments are the
only important ones when convoluting a transverse momentum spectrum
which falls off like $d\hat\sigma_b/d\hat p_T \simeq A/\hat p_T^5$, as
is the case in hadronic collisions. We have in fact that the
convolution of such a perturbative spectrum with a non-perturbative
fragmentation function gives 
\begin{eqnarray*}
\frac{d\sigma_B}{dp_T}  =
\frac{d\hat\sigma_b}{d\hat p_T} \otimes D_{b\to B}(z)  \simeq  
A \int\frac{dz}{z} \left(\frac{z}{p_T}\right)^5 D_{b\to B}(z) = 
\frac{d\hat\sigma_b}{dp_T} D^5_{b\to B} \nonumber
\end{eqnarray*} 
i.e. the spectrum for $B$
hadrons is given by the spectrum for $b$ quarks multiplied by the fifth
moment of the non-perturbative fragmentation function.} These
non-perturbative moments are then used together with a calculation
having the same perturbative features, FONLL~\cite{Cacciari:1998it} 
(Fixed Order plus Next-to-Leading Log - in this case
$\log(p_T^2/m_b^2)$), to evaluate  the cross sections in $p\bar p$
collisions.

The expectation is then that total cross sections be reproduced by the
NLO calculations for $b$ quarks, and that differential distributions for
$B$ hadrons be correctly described by a proper convolution of the FONLL
perturbative spectrum for $b$ quarks and  the non-perturbative
information extracted from LEP data. Notice that a {\sl minimalist} use
of non-perturbative information is made: there is no attempt to fully
describe the hadronization process. Only the relevant phenomenological
information is determined from data and used in the predictions.

A successful comparison will see data and theory in agreement
within their {\sl combined} uncertainties. The theoretical ones
will be assessed by varying as extensively as reasonable the parameters
and the unphysical scales entering the predictions. As for the
experimental errors, it is perhaps worth reminding that only
1-sigma errors are usually shown on the plots, so that non-overlapping
bands do not necessarily point to a solid disagreement.

\section{The Data}

\subsection{$\gamma\gamma$}

\begin{figure}
\begin{center}
\epsfig{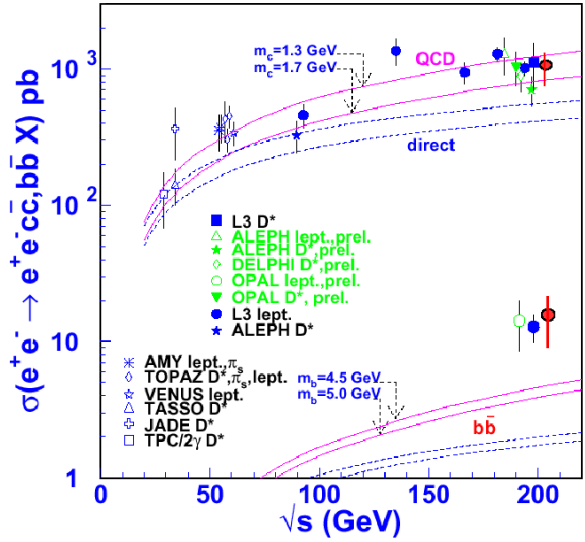}
\caption{\small\label{figgg} Total cross sections for $\gamma\gamma\to c\bar c$ and
$b\bar b$.}
\end{center}
\end{figure}

Let us first consider the gamma-gamma 
data~\cite{Acciarri:2000kd,opal-gg,delphi-gg}, measured in $e^+e^-$
collisions at $\sqrt{S} = 194$~GeV. Unfortunately only the L3 paper is
published in final form.  The results from the three experiments appear
fully compatible, reading $\sigma(e^+e^-\to e^+e^-b\bar bX) = 12.8 \pm
1.7 \pm 2.3$~pb (L3), $14.2 \pm 2.5 ~^{-4.8}_{+5.3}$~pb (OPAL) and
$14.9 \pm 3.3 \pm 3.4$~pb (DELPHI). These results should be compared to
a theory prediction that the experimental papers estimate of the order
of 4 pb, apparently with an uncertainty inferior to 10-15\%.
Figure~\ref{figgg} shows the comparison between theory and experiment
as usually presented at Conferences. 

Four comments are in order here. The first is that the theoretical
uncertainty has probably not been fully explored, missing at least the investigation of
the effect of different photon parton distribution function sets, which are
only approximately known, and the exploration of independent
renormalization and factorization scale variations. Ref.
~\cite{Frixione:2000kt} for instance, while not performing a full exploration
of all variables, still estimates a larger uncertainty.  The second is that,
especially when including theoretical errors, the real distance between data
and theory probably does not exceed 2 sigmas. Hence, claims of data being
``{\it ...in excess of the QCD prediction by a factor of
three}''\cite{Acciarri:2000kd} appear premature, not being backed by adequate
significance. An `excess factor' may correctly refer to central values, but
it should not be used to measure  a `discrepancy' until errors are also
included. The third comment is that
the three experiments essentially relied on the same technique and the same
tool (the PYTHIA Monte Carlo) for extracting the $b$ signal. Hence, their
results are probably strongly correlated and their accord is probably less
significant than it appears. Finally, last but not least, the total cross
section is not what is actually measured by the experiments. Rather, only a
fairly tiny fraction of the cross section is measured with the use of PYTHIA,
and by means of the same Monte Carlo this cross section is then extrapolated to
the full phase space. Needless to say, such a procedure might introduce a
bias, as the theoretical prediction of PYTHIA outside the measured
region cannot of course be considered a priori correct. Unless the uncertainty
that the extrapolation introduces can be reliably estimated, the total cross
section results cannot be considered as real `measurements'. Comparisons of
these data with theoretical predictions should therefore be taken with a grain
of salt.

\subsection{$\gamma p$ and $ep$}

Bottom production cross sections have also been measured in both
photoproduction and Deep Inelastic Scattering (DIS) at HERA by the H1
and ZEUS Collaborations.  The first measurement was performed by the
H1  Collaboration~\cite{Adloff:1999nr} in 1999. While allowing for
experimental and theoretical uncertainties of the order of 10-20\%,
this paper (and in particular the erratum) presented a cross section
(deconvoluted and extrapolated) about a factor of three larger than the
NLO prediction. Shortly thereafter the ZEUS Collaboration also
published its first data on bottom
photoproduction~\cite{Breitweg:2000nz}. Comparison with NLO QCD was
again  performed after deconvolution to parton level and extrapolation
by Monte Carlo  to a more inclusive cross section. While the central
value of the cross section was found to be about a factor of two larger
than the central NLO prediction, large experimental and theoretical
uncertainties should prevent one from inferring an incompatibility between the
two numbers, which should actually be considered compatible at the
1-sigma level. Nevertheless, the paper concludes with the observation
that the result is ``{\it ...consistent with the general observation
that NLO QCD calculations underestimates beauty production...}''.

\begin{figure}
\begin{center}
\epsfig{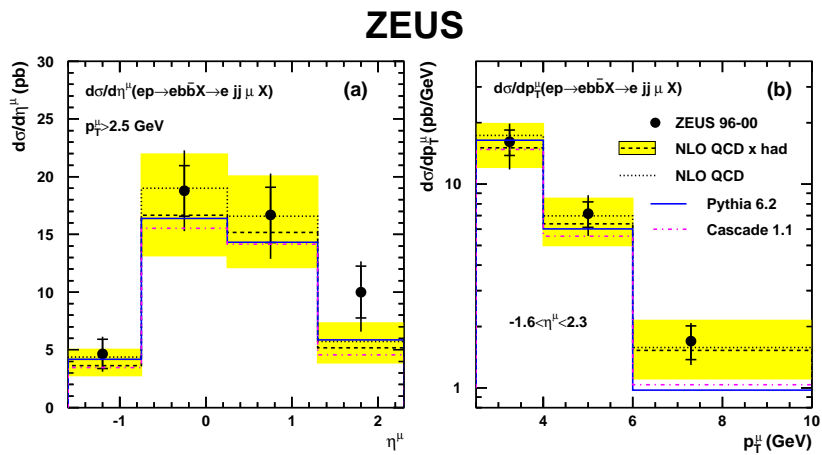}
\caption{\small\label{figzeus} One of the plots 
published in~\protect\cite{Chekanov:2003si} of the observable cross
sections measured by ZEUS and their comparisons to theoretical
predictions.}
\end{center}
\end{figure}

The reliability of these two results being somewhat diminished by the
deconvolutions and/or extrapolations performed by Monte Carlo before
actual comparisons to NLO QCD predictions, the ZEUS Collaboration set
out to perform direct comparisons between theoretical predictions and
real measurements. Photoproduction data~\cite{Chekanov:2003si} are
found to be in very good agreement with NLO QCD (see
figure~\ref{figzeus}), prompting the conclusion that the large excess
found by H1 is not confirmed. It is worth noting that the ZEUS
Collaboration also mentions that the new results are consistent with
their old one~\cite{Breitweg:2000nz}. While this is certainly true from
the statistical point of view, it becomes then apparent that different
conclusions (`above theory' vs. `consistent with theory') were drawn
from compatible measurements. One cannot help noticing that -
apparently - a different weight is given to `sigmas' when comparing
experiment to experiment rather than experiment to theory.

The ZEUS Collaboration has very recently extended the photoproduction analysis
to the DIS regime~\cite{Chekanov:2004tk}. The data are found to be consistent
with NLO QCD predictions, and at most two sigmas higher in a few bins of the
distributions.

\subsection{$p\bar p$}

\begin{figure}
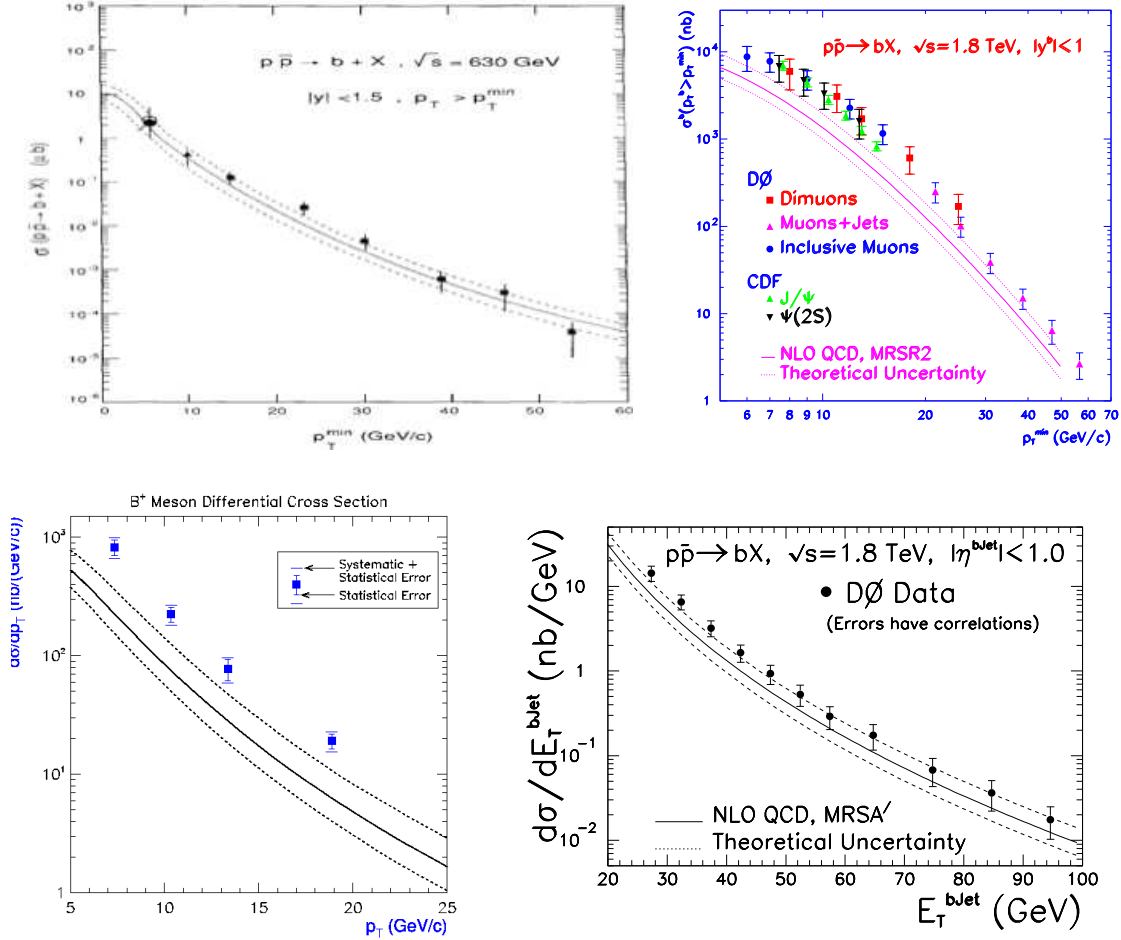

\epsfig{file=ua1.seps,height=6cm,clip=}
~~~~
\epsfig{file=bqplot.seps,height=6.5cm,clip=}\\[.5truecm]
\epsfig{file=diff25_no_descr.seps,height=6cm,clip=}
~~~~
\epsfig{file=bjets.seps,height=6cm,clip=}
\caption{\small\label{poster} A collection of bottom quark production
measurements and comparisons with theory. From top to bottom, left to
right: a) the UA1 results~\cite{Albajar:1990zu}, 
b) a collection of CDF and D0 data, 
c) the CDF Run~I $B^+$ cross section~\cite{Acosta:2001rz}, and 
d) the D0 $b$-jets cross section~\cite{Abbott:2000iv}.
}
\end{figure}

Hadronic collisions were historically the first to produce bottom production
data in collider physics. The first results date back to 1986, and were
published by the UA1 Collaboration, using $p\bar p$ collisions at $\sqrt{S} =
546$ and 630 GeV. In a subsequent paper~\cite{Albajar:1990zu}, the UA1
Collaboration compared bottom quark transverse momentum distributions to the
then recently completed NLO 
prediction~\cite{Nason:1987xz,Nason:1989zy,Beenakker:1988bq,Beenakker:1990ma},
and found a good agreement (see figure~\ref{poster}a). It is worth noting that, while comparing data
deconvoluted to the unphysical quark level, the paper also included data for
the observed $B$ mesons. While a comparison with theoretical predictions at the
hadron level was not performed at that time, it is however still possible now,
thanks to the preservation and publication of the real data. Such a comparison,
performed with the modern tools and framework described in
Section~\ref{sec:paradigm},  shows an agreement similar to the one observed at
the quark level.

The main source of bottom data in $p\bar p$ collisions in recent years
has  been the Fermilab Tevatron, running at $\sqrt{S} = 1800$~GeV first
and 1960 GeV later. The first results from this machine were given
in 1992 by the CDF Collaboration~\cite{Abe:1992fc}. The inclusive cross
section (integrated above a minimum $p_T$ of 11.5 GeV, and within the
central rapidity region $|y|<1$) was published deconvoluted to the quark
level. Its central value was found to be a factor of six larger than
the NLO prediction, but the very large errors only made it a 1.6 sigma
distance and therefore not a significant one.

One year later CDF started publishing~\cite{Abe:1993sj,Abe:1993hr} the
plot of the transverse momentum distribution which will then become the
icon of the supposed `excess' (see figure~\ref{poster}b). The data were
published only at the unphysical quark level and, while reporting
differences of the order of 1-2 sigmas between data and theory, the
papers still conclude that ``...the next-to-leading order
QCD calculation tends to underestimate the inclusive $b$-quark cross
section'', hinting therefore for the first time that a discrepancy might
be present. The same conclusion was reached in a subsequent
paper~\cite{Abe:1995dv}, where cross sections for real particles, $B$
mesons, were finally published. In this case the `disagreement' was
quantified for the first time, a fit to an  overall scale factor
between data and theory yielding 1.9 $\pm$ 0.2 $\pm$ 0.2.

Around the same time CDF published the first data on bottom quarks, the D0
Collaboration also released some preliminary results which were presented at a
number of conferences. Somewhat at odds with the CDF ones, they were in very
good agreement with the NLO predictions. Therefore, weighing the data from both
collaborations, speakers at the conferences (see e.g.
\cite{Zieminski:qv,Bazizi:1994sp}) generally reported a good agreement between
bottom quark data and NLO QCD.

Given this state of affairs, the final D0 data must have caused some surprise
when, published in final form about one year later~\cite{Abachi:1994kj}, they
became more CDF-like, now lying around the upper edge of the theoretical
uncertainty band. The prediction of NLO QCD, however,  was still
considered to be giving an adequate description of the data.

The subsequent pair of D0 papers on this
subject~\cite{Abbott:1999wu,Abbott:1999se}  should have caused  an even
larger surprise. Despite the conclusions of the previous paper
(``adequate description''~\cite{Abachi:1994kj}), in the Introduction
of~\cite{Abbott:1999wu} the previously measured $b$ quark cross section
is now  considered to have been found ``systematically larger'' than
the central value of NLO QCD predictions. This helps to digest the news
that the data now show a considerable excess: ``{\it The ratio of the
data to the central NLO QCD prediction is approximately
three...}''.~\cite{Abbott:1999wu} This strong statement is even upped
in the following paper which, already in the Abstract, 
states that ``{\it We find that next-to-leading-order QCD calculations
underestimate $b$-quark production by a factor of 4 in the forward
rapidity region}''.\cite{Abbott:1999se} 

If we are to take such statements\footnote{In passing, we note that no
uncertainties are explicitly included, making these statements somewhat
void of significance. Where errors are considered, often buried deep in
the text and tables in the papers, they usually neglect the theoretical
uncertainties, leading therefore to an overestimation of the
significance of a possible gap.} at their face value, we are of course
to conclude that by the year 2000 the `excess' appeared firmly
established. Even new physics, in the form of light supersymmetric
particles~\cite{Berger:2000mp}, was
advocated in order to try explaining the discrepancy (see however also
\cite{Janot:2004cy}, which excludes this model using $e^+e^-$ hadronic
data). More conventional
explanations involve trying to consider small-$x$ resummation
effects~\cite{Jung:2003wu}. These results hint that important contributions 
might be present, and certainly deserve further investigations.
However, for the time being they lack overall NLO accuracy, suffer from
large normalization uncertainties, and do not satisfy the requirements
of our `paradigm'. We shall therefore refrain from performing detailed
comparisons using these results.

Two further experimental papers were then added to the field: both are
worth examining closely, albeit for different reasons.
The D0 Collaboration performed a measurement~\cite{Abbott:2000iv} of the
transverse energy distribution of jets containing a bottom quarks, the so
called $b$-jets. These objects are real observables, while at the same time
being largely independent of the fragmentation properties of the bottom quarks.
By contrast, other observables like the $B$ mesons do instead of course depend
on the way the bottom quarks fragment into the bottomed hadrons. D0 found that
the cross section for $b$-jets was compatible with the NLO QCD
prediction~\cite{Frixione:1996nh} (see figure~\ref{poster}d).
The CDF Collaboration updated instead~\cite{Acosta:2001rz} its
measurement for the $B$ mesons transverse momentum distribution, superseding
\cite{Abe:1995dv} and representing  the final analysis for bottom production
with the Tevatron Run~I data: ``{\it The differential cross section is measured
to be $2.9 \pm 0.2\; (\mathrm{stat}\oplus\mathrm{syst}_{p_T}) \pm 0.4\;
(\mathrm{syst}_{fc})$ times higher than NLO QCD predictions...}''
(figure~\ref{poster}c). A couple of
comments are worth making. The first is that, once more,  the errors on this
ratio do not include the theoretical uncertainty, which from perturbative
sources alone would be at least of the order of 20-30\% of the NLO result. 
The second is that the `NLO QCD' prediction must of course include the
non-perturbative information needed to fragment the bottom quark into the $B$
meson. In this experimental paper this fragmentation was performed using the
Peterson et al.~\cite{Peterson:1982ak} functional form, with its free parameter
set to $\epsilon_b = 0.006$, a standard choice dating back to determinations
performed in 1987 by Chrin~\cite{Chrin:1987yd}. Such a procedure however
neglects the notion that neither the bottom quark nor its fragmentation into
$B$ hadrons are physical observables. Neither of them is separately measurable,
only their final combination is. It is therefore wrong in principle (and also,
as we will see, in practice) to rely on a fixed and standard determination of
the non-perturbative fragmentation function and to convolute it with whatever
perturbative calculation is being used. The non-perturbative fragmentation must
rather be determined from data (usually from $e^+e^-$ collisions) using the
same perturbative framework and parameters (bottom mass, strong coupling) of
the calculation which will then be employed to calculate bottom quark
production in $p\bar p$ collisions.

The problem of using the `wrong' non-perturbative fragmentation function can
become irremediable if only the data for the deconvoluted data for the
unphysical $b$ quarks are finally published. These data might in fact be biased
by the deconvolution, but it would be very hard to reconstruct the originally
measured ones. On the other hand, if the real measurements are (also) published
($B$ mesons, $b$-jets, muons or $J/\Psi$'s from $B$ mesons decays), they can be
directly compared with theoretical predictions which include also the
non-perturbative and decay stages.

\begin{figure}
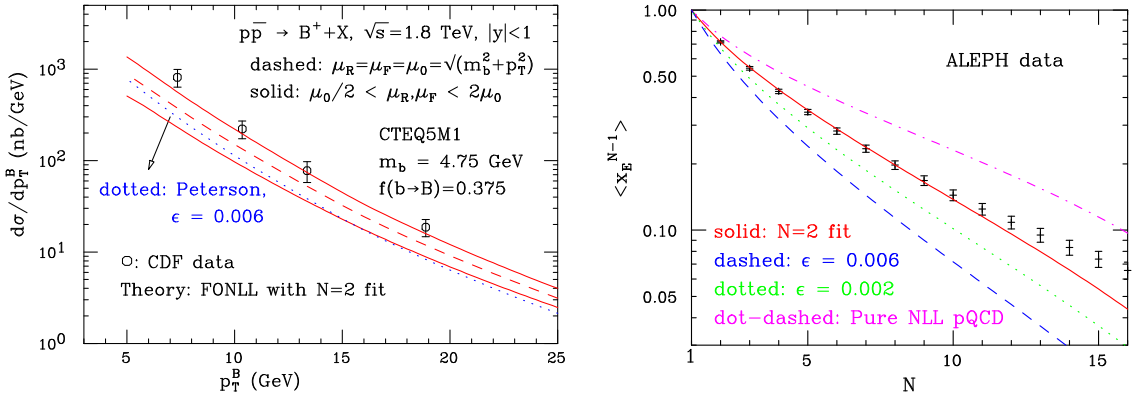

\begin{center}
\epsfig{file=B-hadrons.seps,height=5.2truecm,clip=}
~~
\epsfig{file= cfr-kart-pet.seps,height=5.2truecm,clip=}
\caption{\small\label{cn2002} On the left (a), $B^+$ mesons production at the 
Tevatron and
the new theoretical prediction. On the right
(b), description of moments of $B$ fragmentation data at LEP with different
non-perturbative fragmentation functions. Both plots from
~\protect\cite{Cacciari:2002pa}. }
\end{center}
\end{figure}

This is precisely what was done in~\cite{Cacciari:2002pa}. The non-perturbative
information was determined, as described above, from LEP data in
moment space, and employed - within a consistent perturbative framework -  to
predict the $B$ mesons transverse momentum distribution at the Tevatron.
Figure~\ref{cn2002}a shows how the data from~\cite{Acosta:2001rz} are
compatible with the updated theoretical prediction: the ratio can now be
roughly estimated to be $1.7 \pm 0.5\; (\mathrm{expt}) \pm 0.5\;
(\mathrm{th})$. There is therefore no significant discrepancy between the data
and the theory. Figure~\ref{cn2002}b shows how using the Peterson et al. form
with $\epsilon_b = 0.006$ does indeed underestimate the moments around $N=5$,
consequently leading to an underestimation of the rate in $p\bar p$ collisions.

\section{The Recent Comparisons}

The understanding of the potentially large biases related to the description 
of the non-perturbative fragmentation phase, and the proper inclusion of
uncertainties from all the sources, allowed one to conclude that no significant
discrepancies were probably present in the bottom data collected at the Tevatron
during the Run~I. These data were however always above a minimum $p_T$ of about
5 GeV. Since a harder or softer non-perturbative fragmentation function will
leave the total cross section unchanged while shifting contributions to larger
or smaller $p_T$ values respectively,  it was still possible that such a shift
was only faking a larger cross section. Small-$p_T$ data, and possibly a total
cross section measurement, are therefore crucial for establishing whether
the NLO QCD prediction does indeed account (or not) for the number of bottom
quarks produced at the Tevatron.

Such data, from the Tevatron Run~II, have been recently made
public in preliminary form by the CDF Collaboration~\cite{cdfrun2}, and
promptly compared~\cite{Cacciari:2003uh} to the predictions given by
the framework put forward in~\cite{Cacciari:2002pa}. The data are in the
form of $J/\psi$'s coming from bottomed hadrons $H_b$. The theoretical 
predictions
depend solely on the following calculations and parameters:
\begin{itemize}
\item[$\bullet$] Perturbative inputs
\begin{itemize}
\item FONLL calculation (i.e. full massive NLO calculation plus matching
to NLL resummation), both for $e^+e^-$~\cite{Mele:1990cw} and for 
$p\bar p$~\cite{Cacciari:1998it} collisions
\item bottom quark pole mass $m_b = 4.75$~GeV (varied between 4.5 and 5 GeV)
\item strong coupling ($\Lambda^{(5)} = 0.226$~GeV, i.e. $\alpha_s(M_Z)
= 0.118$)
\item renormalization and factorization scales (varied between
$\mu_0/2 \le \mu_{R,F} \le 2\mu_0$, with $1/2 \le \mu_R/\mu_F \le 2$ and
$\mu_0 \equiv \sqrt{m_b^2+p_T^2}$
\end{itemize}
\item[$\bullet$] Non-perturbative/phenomenological  inputs
\begin{itemize}
\item gluon and light quarks PDFs (CTEQ6M~\cite{Pumplin:2002vw} default 
choice, MRST~\cite{Martin:2002aw} and
Alekhin~\cite{Alekhin:2002fv} sets also used)
\item $b$ quark to $H_b$ hadron fragmentation (fitted to moments of LEP
data, see \cite{Cacciari:2002pa,Cacciari:2003uh})
\item $H_b$ to $J/\psi$ branching ratio, 1.15\%~\cite{Hagiwara:fs} and
decay spectrum (from CLEO~\cite{Anderson:2002jf} or BaBar~\cite{Aubert:2002hc} 
Collaborations)
\end{itemize}
\end{itemize}
After extensive exploration of all the numerically meaningful
uncertainties, the predictions compare to the measured total cross
sections as follows:
\begin{center}
\renewcommand{\arraystretch}{1.2} 
\begin{tabular}{l|c|c} 
& CDF & Theory (FONLL) \\
\hline
$\ss \sigma(H_b\to J/\psi,
p_T(J/\psi)>1.25, \vert y_{J/\psi} \vert<0.6) 
\times {\rm BR}(J/\psi\to\mu^+\mu^-) $
&
$19.9 \; {\ss +3.8 \atop \ss -3.2}_{stat+syst} \; \mbox{nb}$
&
$18.3 \; {\ss +8.1 \atop \ss -5.7}\; \mbox{nb}$
\\
$\ss \sigma(H_b, \vert y_{H_b} \vert<0.6) \times {\rm BR}(H_b\to
J/\psi\to\mu^+\mu^-)$
&
$24.5 \; {\ss +4.7 \atop \ss -3.9}_{stat+syst} \; \mbox{nb}$
&
$22.9 \;  {\ss +10.6 \atop \ss -7.8} \mbox{nb}$
\\
$\ss \sigma(b,\vert y_b \vert < 1) $
&
$29.4 \; {\ss +6.2 \atop \ss -5.4}_{stat+syst} \; \mu\mbox{b}$
&
$25.0 \; {\ss +12.6 \atop \ss -8.1} \; \mu\mbox{b}$\\
\end{tabular}
\end{center}

\begin{figure}
\begin{center}
\epsfig{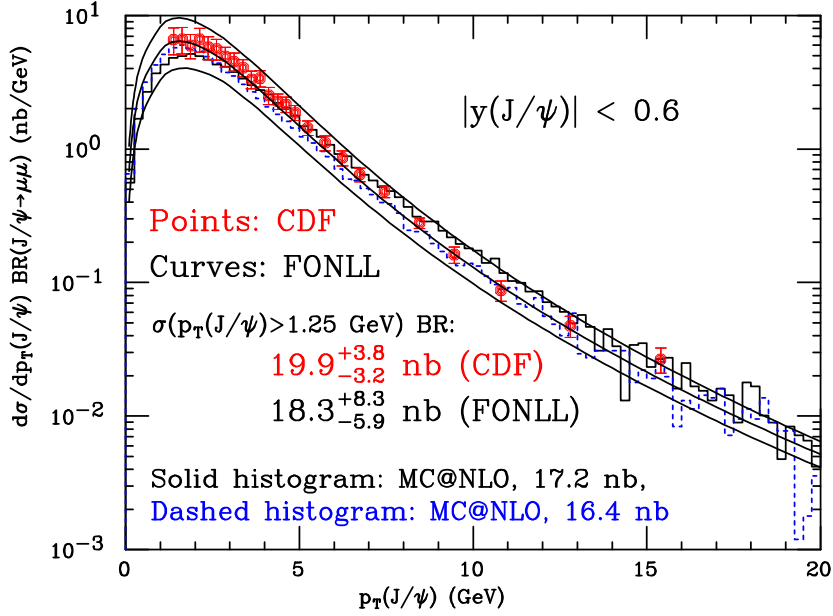}
\caption{\small\label{cdftheory} CDF $J/\psi$ spectrum from $H_b$
decays, compared to theoretical predictions~\protect\cite{Cacciari:2003uh}.}
\end{center}
\end{figure}

The first two lines refer to physical cross sections, measured (and
predicted) in the given visible region. The third line represents the
deconvolution to the quark level. These results clearly indicate full
consistency between theory and experiment within the uncertainties. The
transverse momentum spectrum of the $J/\psi$'s from $b$'s, shown in
figure~\ref{cdftheory}, is equally well described. The reason why the
agreement now looks better than it did in figure~\ref{cn2002} is
twofold. On one hand, the theoretical prediction is increased by
10-20\% by employing a more modern PDF set (CTEQ6M vs. CTEQ5M). On the
other hand, the experimental data, which should have been about 10\% higher 
due to the larger Run~II energy, are instead about 25\% lower (but
still compatible with the old ones within the uncertainties).

\section{Conclusions}

Next-to-leading order QCD appears to be doing a good job in predicting total
and single inclusive  bottom quark production cross sections at the Tevatron
and HERA. Comparisons performed at the {\sl observed hadron} level, rather than
at the unphysical {\sl quark} level, do not seem to show significant
discrepancies. Tevatron Run~II preliminary results are even in very good
agreement. We argue that discrepancies pointed out in the past were either not
very significant, in that the real size of the uncertainties might have been
underestimated or simply overlooked, or that the `data' might have been tainted
by excessive use of Monte Carlo simulation in their extraction, deconvolution
to parton level (perhaps with the wrong fragmentation function), and 
extrapolation to full phase space. An extension of the number of comparisons
performed at the observed particle level and in the visible regions will
certainly help shed light on this point.

New physics has been advocated at some point in order to explain the
presumed discrepancy. While there is of course still room for it within
the uncertainties, at the level of about 30\% in the case of the 
Tevatron data, we argue that its presence is not needed in order to
explain the single inclusive bottom production data.

Finally, we wish to point out that much of the progress done in the
last couple of years has been permitted by the possibility of comparing
theoretical predictions to real data, rather than to
deconvoluted/extrapolated ones. This is not always the case, as
sometimes the original data are not published and are lost forever. We
invite therefore the experimental Collaborations to always publish also
results which do not depend (or depend as little as possible) on
theoretical prejudices (e.g. in the form of a Monte Carlo code) for
their extraction. This will avoid the risk of biasing
them, and will leave open the possibility of performing updated
comparisons in the future.

\acknowledgments{Many thanks go to my collaborators, to G\"unther
Dissertori and Gavin Salam for the careful reading of the manuscript, to  the
experimentalists  I
talked to trying to understand the details of experimental
measurements performed even years in the past, to all the Organizers for
the invitations, and to those who
raised their hands after my presentations at the Conferences, pointing
out what papers I had overlooked or misinterpreted, and offering their
opinion and advice. I have probably not convinced everybody with my
arguments, nor perhaps retained all suggestions, 
but I have certainly enjoyed every conversation.}

\end{document}